\documentclass[aps,twocolumn,showpacs,floatfix,superscriptaddress]{revtex4}
\usepackage{graphicx}
\usepackage{amsmath}
\usepackage{amsfonts}
\usepackage{amssymb}
\usepackage{epsfig}

\begin{document}

\newcommand{\la}{\langle}
\newcommand{\ra}{\rangle}
\def\be{\begin{equation}}
\def\ee{\end{equation}}
\def\bea{\begin{eqnarray}}
\def\eea{\end{eqnarray}}
\def\bma{\begin{mathletters}}
\def\ema{\end{mathletters}}
\newcommand{\one}{\mbox{$1 \hspace{-1.0mm}  {\bf l}$}}
\newcommand{\eins}{\mbox{$1 \hspace{-1.0mm}  {\bf l}$}}
\def\C{\hbox{$\mit I$\kern-.7em$\mit C$}}
\newcommand{\tr}{{\rm tr}}
\newcommand{\half}{\mbox{$\textstyle \frac{1}{2}$}}
\newcommand{\shalf}{\mbox{$\textstyle \frac{1}{\sqrt{2}}$}}
\newcommand{\ket}[1]{ | \, #1  \rangle}
\newcommand{\bra}[1]{ \langle #1 \,  |}
\newcommand{\proj}[1]{\ket{#1}\bra{#1}}
\newcommand{\kb}[2]{\ket{#1}\bra{#2}}
\newcommand{\bk}[2]{\langle \, #1 | \, #2 \rangle}
\def\II{I(\{p_k\},\{\rho_k\})}
\def\ss{{\cal K}}
\tolerance = 10000
\newcommand{\eps}\varepsilon

% You should use BibTeX and revtex.bst for references

%\bibliographystyle{prsty}
%\bibliographystyle{unsrt}

%Title of paper

\title{Cooling toolbox for atoms in optical lattices}
\author{M. Popp}
\address{Max--Planck Institute for Quantum Optics, Garching, Germany}
\author{J.-J. Garcia-Ripoll}
\address{Max--Planck Institute for Quantum Optics, Garching, Germany}
\author{K. G. H. Vollbrecht}
\address{Max--Planck Institute for Quantum Optics, Garching, Germany}
\author{J. I. Cirac}
\address{Max--Planck Institute for Quantum Optics, Garching, Germany}

\begin{abstract}
  We propose and analyze several schemes for cooling bosonic and fermionic
  atoms in an optical lattice potential close to the ground state of the
  no--tunnelling regime. Some of the protocols rely on the concept of
  algorithmic cooling, which combines occupation number filtering with ideas
  from ensemble quantum computation. We also design algorithms that create
  an ensemble of defect-free quantum registers. We study the efficiency of
  our protocols for realistic temperatures and in the presence of a harmonic
  confinement. We also propose an incoherent physical implementation of
  filtering which can be operated in a continuous way.

\end{abstract}

\date{\today}
 \maketitle

\section{Introduction}
\label{Intro}

Ultracold atoms stored in optical lattices can be
controlled and manipulated with a very high degree of
precision and flexibility. This places them among the most
promising candidates for implementing quantum computations
\cite{OL_ent,L_OL_QC_03,W_OL_QC_02,Bloch03,VC04} and
quantum simulations of certain classes of quantum
many--body systems
\cite{CZ_OL_review,CZ_OL_review_1,Fermi-Hubbard,OLspin,OLspin2,JuanjoAKLT,Kagome,ringexchange,ZollerPM,LukinPM}.
Quantum simulation would allow us to understand physical
properties of certain materials at low temperatures that so
far have eluded a theoretical description or numerical
simulation. However, both quantum simulation and quantum
computation with this system face a crucial problem: the
temperature in current experiments is too high. In this
paper we propose and analyze several methods to decrease
the temperature of atoms in optical lattices and thus to
reach the interesting regimes in quantum simulations, as
well as to prepare defect--free registers for quantum
computation.

So far, several experimental groups have been able to load
bosonic or fermionic atoms in optical lattices and reach
the strong interaction regime
\cite{Bloch02,Tonks,ETH_Tonks,ETH_Tonks_1,ETH_Fermi,NIST_OL_04,MIT_OL_Bose,Texas_OL05,Grimm_OL_06}.
The analysis of experiments in the Tonks gas regime
indicates a temperature of the order of the width of the
lowest Bloch band \cite{Tonks}, and for a Mott Insulator
(MI) a temperature of the order of the on-site interaction
energy has been reported \cite{temp_Mott,ETH_Tonks}. For
fermions one observes temperatures of the order of the
Fermi energy \cite{temp_Fermi,
  temp_Fermi_ETH, ETH_Fermi_mol}. Those temperatures severely restrict the
physical phenomena that can be observed and the quantum
information tasks that can be carried out with these
lattices.

One may think of several ways of cooling atoms in optical
lattices. Since the process of loading atoms into the
lattice may lead to additional heating \cite{Bloch06_adiab}
we focus here on methods that operate once the lattice
potential has been raised. For example, one may
sympathetically cool the atoms in the lattice using a
second Bose--Einstein condensate
\cite{Fermi-Hubbard,DFZ04}. A completely different approach
is the \emph{filtering} scheme of Rabl \emph{et al.}
\cite{RZ03}. It operates in the no-tunnelling regime,
transferring atoms between optical lattices so as to create
a configuration with one atom per site. Such a loading
scheme can, however, originate holes due to imperfections
in the original cloud.  Our analysis of filtering in the
presence of a harmonic trap has shown that these holes,
which are preferably located at the borders of the cloud,
result in a considerable amount of entropy \cite{Cool1}.

Here we propose and investigate several cooling schemes
which overcome the limitations of filtering. The first set
of schemes uses discrete operations to make atoms in
different sites interact, thus concentrating the entropy on
some atoms which are then expelled from the lattice. Due to
the similarity with quantum information processing, we term
this kind of methods {\em
  algorithmic cooling} \footnote{Note that our concept of \emph{algorithmic cooling of atoms} has
to be clearly distinguished from \emph{algorithmic cooling
of spins} which is a novel technique that allows to create
highly polarized ensembles of spins in the context of NMR
experiments, see e.g. P. O. Boykin, T. Mor, V.
Roychowdhury, F. Vatan, and R. Vrijen, Proc. Natl. Acad.
Sci. USA {\bf 99}, 3388 (2002).}. The second set of cooling
methods combine filtering with either particle hopping or
evaporative cooling techniques. All our cooling protocols
are based on translation invariant operations (i.e. do not
require single--site addressing) and consider the residual
harmonic confinement present in current experiments.
Although we will be mostly analyzing their effects on
bosonic atoms, they can also be trivially generalized to
fermions.

This work has several spin-offs. First of all, we have
designed algorithms that very efficiently remove all
residual defects from a Mott insulator, thus producing an
ensemble of perfect registers for quantum computing
\cite{VC04}. Second, we propose how to create pointer atoms
at the endpoints of a perfect quantum registers and show
how these pointers can be used to tailor the register to a
specific length. Finally, since filtering is an important
ingredient of all our algorithms, we have also designed a
new incoherent version of filtering procedure which is
better adapted to our protocols than the adiabatic scheme
from Ref.~\cite{RZ03}.

The paper is organized as follows. We start in
Sect.~\ref{Init} describing our system in terms of the
Bose-Hubbard model and discussing the properties of current
experimental states, such as densities, temperature and
entropy. We also summarize the basic tools and figures of
merit which underlie our cooling schemes.
Sect.~\ref{Sect_filtering} is devoted to a particular tool,
filtering. We describe the effect of filtering on the
entropy and temperature for realistic initial states and
study the optimal choice of experimental parameters. In the
last part we propose an implementation of filtering which
is similar to a radio-frequency knife. Once we have
described all tools, in Sect.~\ref{Sect_gscool} we
introduce several protocols for cooling to the ground
state. This includes the algorithmic cooling in
Subsect.~\ref{algo-cool} and two other protocols in the
following subsections. Based on similar ideas,
Sect.~\ref{Sect_qrcool} presents a quantum protocol that
produces an ensemble of defect-free quantum registers and
shows how to create pointer atoms on these registers.  We
conclude this work with some remarks concerning possible
variants and combinations of our cooling schemes. In
Appendix A we present a detailed description of the
numerical method that we use to simulate classically
correlated states.

\section{Physical system} \label{Init}

\subsection{Bose-Hubbard model}

We consider a gas of ultra-cold bosonic atoms which have
been loaded into a three dimensional (3D) optical lattice.
This lattice is created by six laser beams of wave vector
$k=2\pi/\lambda$ propagating along three orthogonal
directions. If the laser light is off-resonant with any
atomic transition, the AC Stark effect induces a periodic
potential on the atoms of the form:
\begin{equation}
\label{Vlattice} V(x,y,z)=V_{0x} \sin^2(2k x) + V_{0y}
\sin^2(2k y)+ V_{0z} \sin^2(2k z),
\end{equation}
with a strength or ``lattice depth'' $V_0$ proportional to
the dynamic atomic polarizability and the laser intensity.
The Gaussian profile of the laser beams creates an
additional harmonic confinement which can be compensated by
additional magnetic or optical confining elements
\cite{Tonks,Bloch05}.

In the following we will mostly be concerned with
one-dimensional (1D) lattices. In other words we will
assume that the lattice potential is so strong along two
directions, $V_{0y},V_{0z}\gg V_{0x}$ that tunnelling is
only allowed along the third one. We will also assume that
the confinement along all directions is still much stronger
than the atomic interaction strength. Under these
conditions, the atoms can be described using a single band
Bose-Hubbard model (BHM) \cite{oplat98}, which for a
lattice of length $L$ reads {\small\begin{equation}
\label{BHM}
 H_{BH}=\sum_{k=-L/2}^{L/2} \left[-J (a^\dagger_k a_{k+1} + h.c.)+
  \frac{U}{2} n_k (n_k-1)+ b  k^2 n_k \right] .
\end{equation}}
The parameter $J$ denotes the hopping matrix element
between two adjacent sites, $U$ is the on-site interaction
energy between two atoms and the energy $b$ accounts for
the strength of the harmonic confinement. Second
quantization operators $a_k^\dagger$ and $a_k$ create and
annihilate, respectively, a particle on site $k$, and
$n_k=a^\dagger_k a_k$ is the occupation number operator.
Since the tunnelling rate decreases exponentially with the
trapping strength, $V_0$, while the on-site interaction $U$
remains almost constant \cite{oplat98}, we have adopted
this last value as the unit of energy for our work. Even
more important, this exponential dependence allows one to
reach the strong interaction regime $U \gg J$. At $ U
\approx 11.6 ~J$ the ground state changes from a superfluid
(SF) to a Mott insulator (MI) \cite{oplat98,Bloch02}. In
the SF regime particles are delocalized over all lattice
sites. In 1D without harmonic trap the SF ground state
reads:
\begin{equation} |
\psi_0 \ra_\textrm{SF} = \left(\sum_{k=-L/2}^{L/2-1}
a_k^\dagger \right)^{\otimes N} |0\ra,
\end{equation}
where $N$ is the particle number. In the MI regime
particles are localized at individual lattice sites. For
integer filling factors $\nu=N/L$ the ground state is given
by:
\begin{equation} \label{gs_MI_1} |\psi_0 \ra_\textrm{MI}
= \bigotimes_{k=-L/2}^{L/2-1} \left( a_k^\dagger
\right)^\nu |0\ra.
\end{equation} The presence of a harmonic trap typically
leads to a wedding cake structure of MI states with
different $\nu$. For very shallow traps one obtains only a
single MI phase, which has filling $\nu=1$ and which is
centered around the bottom of the trap. In the Fock basis
with respect to lattice sites this MI state can be written
as: {\small \begin{equation} \label{gs_MI} |\psi_0
\ra_\textrm{MI}= |0_{L/2} \ldots 0_{-N/2-1} 1_{-N/2} \ldots
1_{N/2-1} 0_{N/2} \ldots 0_{L/2-1} \ra.
\end{equation}} This state will be the
target ground state for  our cooling schemes.

\subsection{Initial states} \label{Sect_initstate}

Throughout this paper we will work with 1D thermal states
in the grand canonical ensemble, which are characterized by
two parameters: the temperature $k T = 1/ \beta$ and the
chemical potential $\mu$. We are particularly interested in
the no-tunnelling limit\footnote{The condition for the
no-tunnelling regime in the presence of a harmonic trap is
given by \mbox{$|J/(bN)|^2 \ll 1$}, i.e. the hopping matrix
element is much smaller than the average energy spacing
between single particle states located at the borders of
the trap. If desired one can also demand $|J/b|^2 \ll 1$,
which ensures that single particle eigenfunctions even at
the bottom of the trap are localized on individual lattice
wells \cite{Menotti}.}, $J\rightarrow 0$. In this limit the
Hamiltonian (\ref{BHM}) becomes diagonal in the Fock basis
of independent lattice sites: $\{ |n_{-L/2} \ldots n_0
\ldots n_{L/2-1}\ra \}$ and the density matrix becomes a
product of thermal states for each lattice site,
\begin{equation}
\label{rho_ind} \rho=\frac{1}{\Theta} e^{-\beta (H_{BH}-
\mu
  N)}= \bigotimes_{k=-L/2}^{L/2-1} \rho_k.
\end{equation}
This simplifies calculations considerably and for instance
the von-Neumann entropy can be written as the sum over
single--site entropies
\begin{equation}
\label{vonNeumann} S(\rho)=\tr( \rho \log_2 \rho)=\sum_k
S(\rho_k).
\end{equation}

\begin{figure}[t]
\includegraphics[width=0.7\linewidth]{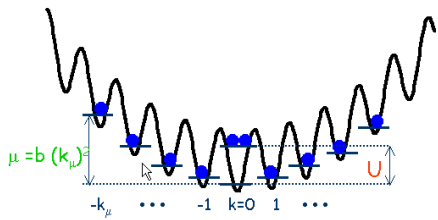}
\caption{Illustration of important energy scales for
bosonic atoms in an
  optical lattice in the no-tunnelling regime: on-site interaction energy
  $U$, chemical potential $\mu$, and harmonic confinement $V_{ho}=b k^2$.
  The characteristic size of the atomic cloud is given by
  $k_\mu=\sqrt{\mu/b}$.}
\label{fig_parameters}
\end{figure}

Let us now study thermal states of the form (\ref{rho_ind})
in more detail. In Fig.~\ref{fig_parameters} we have
depicted graphically all relevant energy scales in the
no-tunnelling limit.  The chemical potential determines the
size of the cloud via the relation $\mu=b k_\mu^2$, where
$k_\mu$ denotes the site at which $\langle n
\rangle_{k_\mu}=0.5$. For $\mu \approx U$ singly occupied
sites at the border of the cloud become energetically
degenerate with doubly occupied sites in the center.

\begin{figure}[t]
\includegraphics[width=0.4\linewidth]{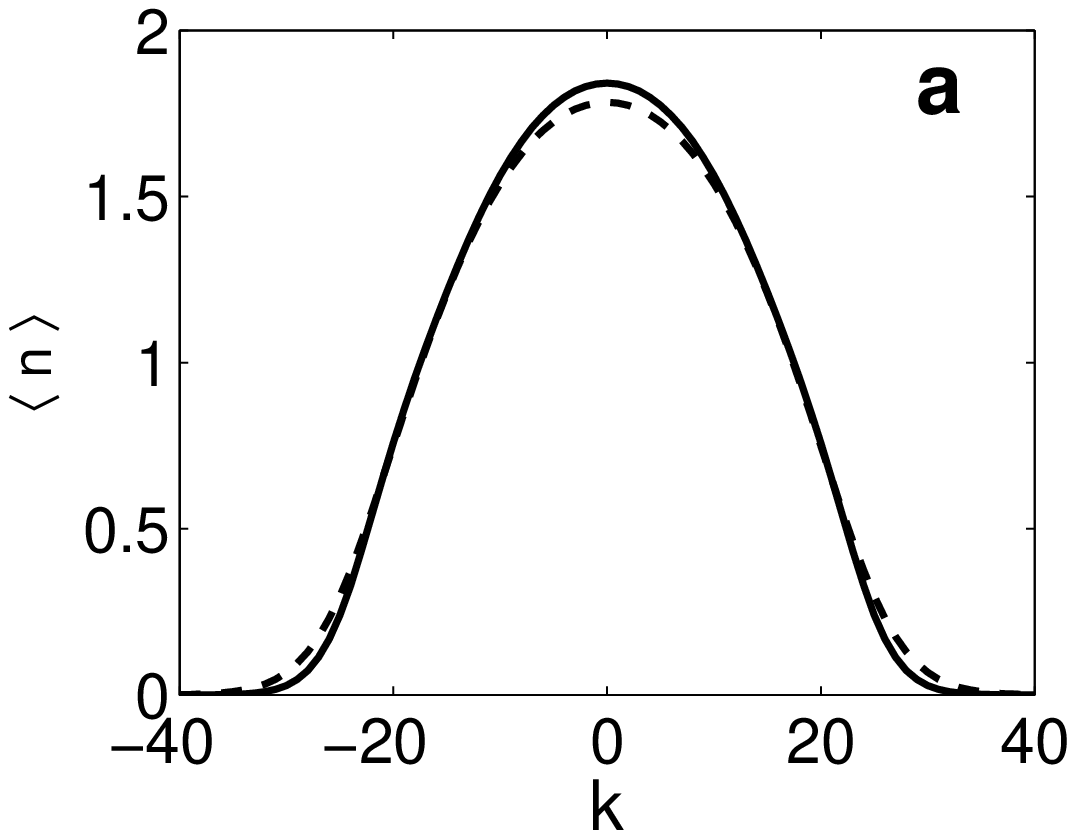}
\includegraphics[width=0.4\linewidth]{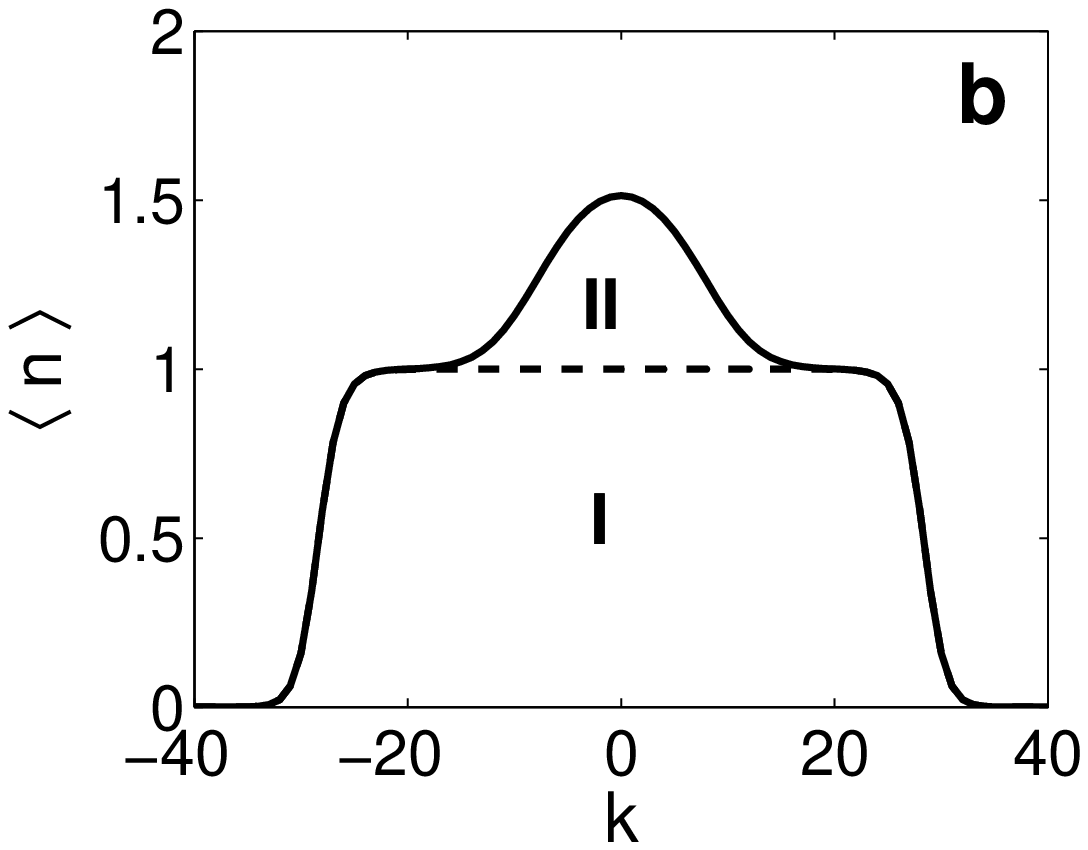}
\caption{(a) Density distribution of two thermal states
with equal R\'enyi
  entropy $S_2/N=0.82$ ($S/N=1$ in MI phase) and equal particle number
  $N=65$. At hopping rate $J/U=0$ (solid) the temperature is given by $ k
  T/U=0.32$ and at $J/U=0.16$ ($V_0=5 E_r$) (dashed) one obtains $ k
  T/U=0.46$. The harmonic confinement is fixed at $U/b=370$ ($V_0=22 E_r$ in
  transverse direction). (b) The separation into two fermionic phases
  becomes clearly visible in the density profile of a thermal state at low
  temperatures. Numerical parameters: $J=0$, $kT/U=0.072$, $\mu/U=1$,
  $U/b=800$, ($N=65$, $s=0.5$).}
\label{fig_initial_1D}
\end{figure}

The analysis of recent experiments in the Mott regime
\cite{Tonks,temp_Mott} reveals a substantial temperature of
the order of the on-site interaction energy $U$. This
result is consistent with our own numerical calculations
and translates into an entropy per particle $s:= S/N
\approx 1$. The particle number in a 1D tube of a 3D
lattice as in \cite{Bloch02} typically ranges between
$N=10$ and $N=130$ particles. A representative density
distribution corresponding to such initial conditions (with
$N=65$) is plotted in Fig.~\ref{fig_initial_1D}a.
In this example the inverse temperature is given by $\beta
U=3.1$.  Since our cooling protocols lead to even lower
temperatures, we will from now on focus on the \emph{low
temperature regime}, $\beta U \gg 1$. Moreover, we will
only consider states with at most two particles per site,
which puts the constraint $\mu \lesssim 2 U- 1 / \beta $ on
the chemical potential. Such a situation can either be
achieved by choosing the harmonic trap shallow enough or by
applying an appropriate filtering operation \cite{RZ03}.

\begin{figure}[t]
\includegraphics[width=0.8\linewidth]{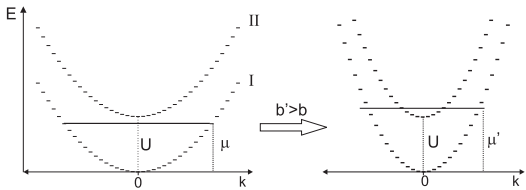}
\caption{Effective description of thermal states in the
no-tunnelling limit
  in terms of independent fermions occupying two energy bands. The
  dispersion relations are $\eps_\textrm{I}=b k^2$ and $\eps_\textrm{II}=b
  k^2 +U$, where $k$ denotes the lattice site and $U$ is the interaction
  energy. Increasing the harmonic trap strength from $b$ to $b'$ increases
  the chemical potential to $\mu'$ so that the population of the upper band
  becomes energetically favorable. In the bosonic picture this process
  corresponds to the formation of doubly occupied sites.}
\label{fig_bands_2}
\end{figure}

Under the assumptions $e^{\beta U} \gg 1$ and $\mu -U/2
\gtrsim b + 1/(2 \beta)$ we have shown in \cite{Cool1} that
the density distribution of the initial state
(\ref{rho_ind}) can be separated into regions that are
completely characterized by fermionic distribution
\begin{equation}
\label{nf} f_k (b, \beta, \mu) = \frac{1}{1+e^{ \beta (b
k^2 -\mu)}}.
\end{equation}
To be more precise, at the borders of atomic cloud, $b k^2
\gg \mu-U/2 +1 /(2 \beta)$, the mean occupation number is
given by $\la n_k \ra \approx n_\textrm{I}(k)$ with
$n_\textrm{I}(k):=f_k(b, \beta, \mu)$, while in the center
of the trap, $b k^2 \ll \mu-U/2 -1 /(2 \beta)$, one has:
$\la n_k \ra \approx 1+ n_\textrm{II}(k)$ with
$n_\textrm{II}(k):=f_k(b, \beta, \mu_\textrm{II})$ and
effective chemical potential $\mu_\textrm{II}:=\mu-U$. Note
also that the underlying MI phase in the center of the trap
is well reproduced by the function $n_\textrm{I}(k)$ ,
which originally has been derived for the border region. As
a consequence, the density distribution for the whole
lattice can be put in the simple form: $\la n_k \ra \approx
n_\textrm{I}(k)+n_\textrm{II}(k)$, which corresponds to two
fermionic phases I and II , sitting on top of each other
[Fig.~\ref{fig_initial_1D}b]. In other words, the initial
state of our system can effectively be described in terms
of non-interacting fermions, which can occupy two different
energy bands I and II, with dispersion relations
$\eps_\textrm{I}=b k^2$ and $\eps_\textrm{II}=b k^2 +U$,
respectively [Fig.~\ref{fig_bands_2}].

So far this fermionic interpretation is just based on the
properties of the density profile of thermal states in the
MI regime. In order to put this fermionic picture on more
general grounds one  can  fermionize the BHM (\ref{BHM})
directly. In \cite{Cool1} we have shown that  the dynamics
at finite $J$ can effectively be described by the following
fermionic Hamiltonian:
\begin{eqnarray}
  \tilde{H} &=& -J\sum_k \left( c^\dagger_k c_{k+1}+ \sqrt{2} c^\dagger_k
    d_{k+1} + 2d^\dagger_k d_{k+1} + \textrm{H.~c.}\right)\nonumber\\
  &+&\sum_k\left [b k^2 c_k^\dagger c_k + (b k^2 + U) d_k^\dagger d_k
  \right].
  \label{H_fermi}
\end{eqnarray}
Here, the fermionic operators $c_k$ and $ d_k$ refer to
energy band I and II, respectively. This effective
description is self-consistent  as long as the probability
of finding a particle--hole pair is negligible, i.e. $\la
c_k c_k^\dagger d_k^\dagger d_k \ra \approx 0$. We have
shown above that this fulfilled for thermal states at low
temperatures and at negligible tunnelling. Later we will
see to what extent the validity can be extended to finite
tunnelling rates.

In general, the state of the system will however exhibit
both classical and quantum correlations. In this  case we
have to represent the state in terms of  a Matrix-Product
State (MPS) \cite{Fannes,RomerPRL}. This concept can also
be used to compute 1D thermal states numerically
\cite{DMRGmixed}. Based on this method one can, for
instance,  estimate how the temperature of a 1D tube
changes when passing from the MI to the SF regime. Assuming
that the process is thermodynamically adiabatic one obtains
the new temperature with the following procedure. We tune
the parameters of a thermal state in the SF regime until
the expectation  values for the entropy and particle number
match the corresponding values of the initial thermal state
in the MI regime. This can be illustrated with the
following example. Starting from a representative state in
the no-tunnelling regime with $s=1$ and $N=65$ we have to
tune the temperature to $k T =0.46U =2.9 J$ at $J/U=0.16$
($V_0=5 E_r$) in order to leave $s$ and $N$ unchanged
[Fig.~\ref{fig_initial_1D}a]. Hence, in the SF regime one
faces a substantial temperature of the order of the width
of the lowest Bloch band. Note, however, that this is only
a lower bound to the true temperature, because our approach
does not include any sort of heating processes induced by
the adiabatic evolution.

\subsection{Entropy as figure of merit}

We will show below that algorithmic protocols are suited
both for ground state cooling and the initialization of
quantum registers. The goal in both cases is to create a
pure state under the constraint of keeping a large number
of particles. Given the fact that our protocols converge to
the desired family of states, we can measure the
performance of the protocol by computing the mixedness of
the state. This mixedness can in turn be quantified using
the von--Neumann entropy $S$ (\ref{vonNeumann}). In some
cases, such as for finite hopping $J$ we will not be able
to compute the von-Neumann entropy efficiently. We will
then refer to the R\'enyi entropy
\begin{equation}
S_2=-\log_2 (\tr \rho^2), \label{Renyi}
\end{equation}
which is a lower bound $S_2 \leq S$ and can be evaluated
using MPS [Appendix A].

In order to assess the efficiency of a protocol in
achieving our objectives we define two figures of merit.
The ratio of the entropies \emph{per} particle after and
before invoking the protocol, $s_f/s_i$, quantifies the
amount of cooling. The ratio of the final and initial
number of particles, $N_f/N_i$, quantifies the particle
loss induced by the protocol.  Note that these figures of
merit can sometimes be misleading and should therefore be
applied with care. In the case of ground state cooling the
entropy is only a good figure of merit if the state of the
system after the cooling protocol, $\rho_f$, is close to
thermal equilibrium. If this is not fulfilled, we use an
effective thermal state, $\rho_f \rightarrow \rho_{eff}$,
with the same number of particles, $N$, and energy, $E$, to
compute the figures of merit. For instance, the final
entropy is given by $S(\rho_{eff})$. Given that our system
can somehow thermalize, these will be indeed the properties
of our final state. Finally, it is important to point out,
that other variables, like energy or temperature, are not
very well suited as figures of merit, because they depend
crucially on external system parameters like the trap
strength.

Note also that in the case of quantum registers it can be
erroneous to assume that a finite value of the final
entropy implies the existence of defects. For example, we
will propose a protocol below that generates a state which
his an incoherent superposition of perfect quantum
registers with varying length and position. This state has
some residual entropy but it is ideal for quantum
information processing.

\subsection{Basic operations}\label{Operations}

All our cooling protocols rely on a set of translationally
invariant quantum operations that can be realized in
current experiments with optical lattices. These operations
are:

\textit{(i) Particle transfer:} Depending on the occupation
numbers, an integer number of particles is transferred
between internal states $|a\ra$ and $|b\ra$. This process
can be described by the unitary operation
\begin{equation}
\label{U_nm} U_{m,n}^{m+x,n-x}: |m,n\rangle \leftrightarrow
|m+x,n-x\rangle,
\end{equation}
where $x$ is an integer and $|m,n\rangle$ are Fock states
with $m$ and $n$ atoms in internal states $|a\ra$ and
$|b\ra$, respectively.  Note that  for this unitary
operator it holds $U_{m,n}^{m+x,n-x}=U_{m+x,n-x}^{m,n}$.
Certain operations, like $U_{2,0}^{0,2}$ or
$U_{1,0}^{0,1}$, have already been demonstrated
experimentally in an entanglement interferometer
\cite{WB04}.

\textit{(ii) Lattice shifts:} We denote by $S_x$ the
operations which shift the $|b\ra$ lattice $x$ steps to the
right. For example, $S_{-1}$ transforms the state
$\otimes_{k} |m_k,n_{k}\rangle_k$ into $\otimes_{k}
|m_k,n_{k+1}\rangle_k$. This operation can be realized in
state-dependent lattices by adjusting the intensity and
polarization of the laser beamsa
\cite{OL_ent,L_OL_QC_03,W_OL_QC_02,Bloch03,OL_spin_99}.

\textit{(iii) Merging and splitting of lattice sites:}
Making use of superlattices \cite{super} one can either
merge adjacent lattice wells or split a single site into a
pair of two sites.

\textit{(iv) Emptying sites:} All atoms in internal state
$|b \ra$ are removed from the system.  We denote this
operation by $E_b$.  It transforms the state $\otimes_{k}
|m_k,n_{k}\rangle_k$ into $\otimes_{k} |m_k,0 \rangle_k$.
Experimentally, this can be achieved either by switching
off the lattice potential acting on $|b \ra$ or by coupling
this state resonantly to an untrapped state.

\begin{figure}[t]
\includegraphics[width=0.7\linewidth]{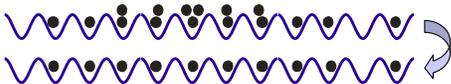}
\caption{Illustration of filter operation $F_1$ which aims
at producing a
  uniform filling of one atom per site. Defects arising from holes cannot be
  corrected.}
\label{fig_P1_lattice}
\end{figure}

\textit{(v) Filtering:} This means particle transfer
operations of the form $U_{m,0}^{M,m-M}$, followed by
$E_b$, for all $m >M$. After tracing out the subsystem
$|b\ra$, the filter operation is described by a completely
positive map acting solely on atoms in state $|a\ra$:
\begin{eqnarray}
\label{Fn} F_M[\rho]: & & \sum_{n,m}
\rho_{n,m}|n\rangle\langle m| \to\\
& &\sum_{n,m\leq M}\rho_{n,m}|n\rangle\langle m| + \sum_{n>
  M}\rho_{n,n} |M\rangle\langle M|.  \nonumber
\end{eqnarray}
The first proposal for a coherent implementation of the
operation $F_1$ [Fig.~\ref{fig_P1_lattice}] appeared in
Ref.~\cite{RZ03}. More recently, a scheme based on resonant
control of interaction driven spin oscillations has been
put forward \cite{Bloch_filter06}. In \cite{Cool1} we have
proposed an ultra-fast coherent implementation of $F_1$
relying on optimal laser control. In this paper we will
propose an incoherent realization of filtering, which has
the special virtue that it can be applied continuously.

\section{Filtering} \label{Sect_filtering}

We now study the filtering operation $F_1$ in
Eq.~(\ref{Fn}) in more detail. This operation is especially
relevant for cooling because it produces a state close to
the ground state of the MI regime. Thus, it can serve as a
benchmark which has to be beaten by alternative cooling
schemes. In this section we begin summarizing the
analytical results about filtering from Ref.~\cite{Cool1}
and discuss the conditions to reach optimal cooling
efficiency. We close this section with a possible
experimental implementation of filtering which is based on
an incoherent coupling of atoms to a continuum of untrapped
states.

\subsection{Theory}

\begin{figure}[t]
\includegraphics[width=0.8\linewidth]{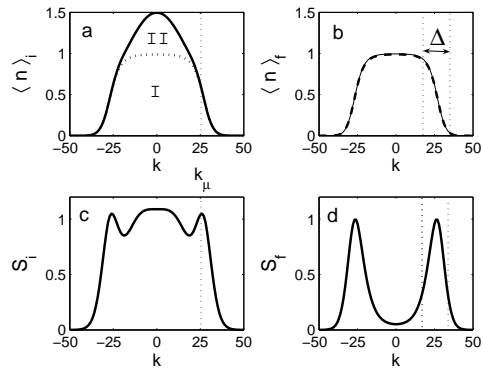}
\caption{Spatial dependence of the mean occupation number
$\la n \ra$ and
  entropy $S$ before (left) and after (right) the application of the filter
  operation $F_1$. The final particle distribution can be well described by
  Eq.~(\ref{nf}) (dashed line). Numerical parameters for initial state:
  $N_i=65$, $s_i$=1, $U/b=700$, $\beta U=4.5$ and $\mu/U=1$. Figures of
  merit: $s_f/s_i=0.56$ and $N_f/N_i=0.80$.} \label{fig_P1}
\end{figure}
In Fig.~\ref{fig_P1} we have depicted the particle and
entropy distributions before and after one filtering step,
$F_1$.  One observes that a nearly perfect MI phase with
filling factor $\nu=1$ is created in the center of the
trap.  Defects in this phase are due to the presence of
holes and concentrate at the borders of the trap. This
behavior is reminiscent of fermions, for which excitations
can only be created within an energy range of order $kT$
around the Fermi level. This numerical observation can
easily be understood with our previous analysis of the
initial state. Filtering removes phase II, which is due to
doubly occupied sites, and leaves the fermionic phase I
unaffected [Fig.~\ref{fig_P1}a].

The fermionic picture allows us to to find a simple
estimate for the final entropy. From the expression for the
density of states, $g(E)=1/\sqrt{b E}$, one immediately
obtains that the number of states (lattice sites) within an
energy range of $ 2 k T$ around the chemical potential
$\mu$ is given by:
\begin{equation}
  \label{Delta} \Delta:= \frac{2}{\beta \sqrt{b \mu}}.
\end{equation}
This parameter is characteristic for the tail width of the
density distribution [Fig.~\ref{fig_P1}] and will in the
following be central for the analysis of our protocols.
Since the final entropy must be localized at these sites we
expect that $S_f\approx \Delta$. Indeed, the rigorous
derivation in \cite{Cool1} yields $S_f= \sigma_\textrm{I}
~\Delta = \sigma_\textrm{I} ~ N_f/\beta \mu$, with
$\sigma_\textrm{I}=\pi^2/(6 \ln 2)$. Here, we have
introduced the final number of particles $N_f$, which can
be well approximated by the parameter $\overline{N}$:
\begin{equation}
  \label{N_bar} \overline{N}=2 \sqrt{\frac{{\mu}}{{b}}}.
\end{equation}
Hence, the final entropy per particle can be written as:
\begin{equation}
  \label{s_f}
  s_f=\frac{S_f}{N_f}\approx \sigma_\textrm{I} \frac{1}{\beta \mu},
\end{equation}
which reflects the size of the region of defects $\Delta$
in units of the system size $\overline{N}$. For the special
choice $\mu=U$ (or equivalently \mbox{$\la n_0 \ra=1.5$})
one finds the following expressions for our figures of
merit \cite{Cool1}:
\begin{eqnarray}
  \frac{s_f}{s_i}&\approx& \frac{\sigma_\textrm{I}}{
    \sqrt{\beta U}} \frac{\eta_\textrm{II}+2 \sqrt{\beta
      U}}{\sigma_\textrm{II} \sqrt{\beta U} +2~\sigma_\textrm{I}}, \label{cool_eff} \\
  \frac{N_f}{N_i} & \approx & \frac{1}{1+ \frac{\eta_\textrm{II}}{2
      \sqrt{\beta U}}},
\end{eqnarray}
with numerical parameters $\sigma_\textrm{II} \approx
2.935$ and $\eta_\textrm{II}=(1-\sqrt{2}) \sqrt{\pi} \zeta
(1/2)\approx 1.063$.  This result shows that filtering
becomes more efficient with decreasing temperature, since
$s_f/s_i \propto 1/\sqrt{\beta U}\rightarrow 0$ and
$N_f/N_i \rightarrow 1$ for $\beta U \rightarrow \infty$.

It is important to note that the state after filtering is
not an equilibrium state, because it is energetically
favorable that particles tunnel from the borders to the
center of the trap, thereby forming doubly occupied sites.
However, the system can easily be brought to equilibrium by
adapting the trap strength. While tunnelling is still
suppressed, one has to decrease the strength of the
harmonic confinement to a new value $b'$, with $b' \leq b ~
U/(2 \mu)$. The system is then in the equilibrium
configuration $f_k(b', \beta', \mu')$ (\ref{nf}) with
rescaled parameters $T'=T~b'/b$ and $\mu'=\mu~b'/b \leq
U/2$.  This observation shows that it is misleading to
infer the cooling efficiency solely from the ratio $T'/T$,
because it depends crucially on the choice of $b'$.

\subsection{Optimal initial conditions}

\begin{figure}[t]
  \includegraphics[width=0.8\linewidth]{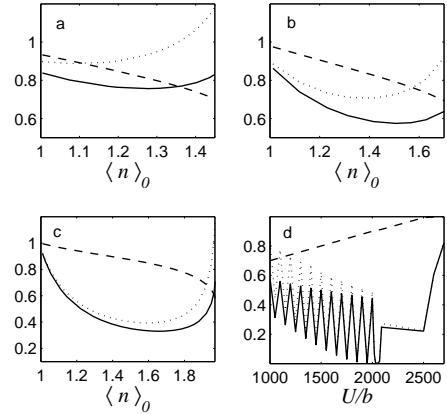}
  \caption{Cooling efficiency of the filtering operation $F_1$ for fixed
    initial entropy $s_i$ and particle number $N_i$ as a function of the
    harmonic confinement. The change of the trap is indicated by either the
    mean central occupation number $\la n_0 \ra$ or the inverse trap
    strength $U/b$. We plot the figures of merit $s_f/s_i$ (solid),
    $N_f/N_i$ (dashed) and the weighted quantity $(s_f N_i)/(s_i N_f)$
    (dotted). The initial parameters are $N_i=100$ and (a) $s_i=1.5$, (b)
    $s_i=1$, (c) $s_i=0.5$ and (d) $s_i=0.05$.} \label{fig_P1_opt}
\end{figure}

Let us now study how the cooling efficiency of filtering
depends on the initial state variables $b$, $\beta$ and
$\mu$. Since the initial temperature is dictated by the
experimental setup, we consider only the trap strength $b$
and the chemical potential $\mu$ (which can be varied via
the particle number $N$) as free parameters. Since our
figures of merit are computed per particle we expect a weak
dependence on $N$ and therefore focus on the $b$
dependence.  This can be studied in terms of the mean
central occupation number $\la n \ra_0$. For instance, in
the special case $\la n_0 \ra=1.5$, we have obtained the
analytical expression (\ref{cool_eff}) for the ratio
$s_f/s_i$, which exhibits a strong temperature dependence.
In contrast, for $\la n_0 \ra=2$ one finds that the cooling
efficiency becomes independent of the temperature: $s_f/s_i
\approx 1/\sqrt{3}$ for $\beta U \gg 1$. This can be
understood from the presence of a $\nu=2$ MI phase in the
center of the trap, which does not contain entropy.  In the
opposite regime $\la n_0 \ra=1$ the protocol has no cooling
effect at all. For general $\la n_0 \ra$, we have computed
the figures of merit numerically exact
[Fig.~\ref{fig_P1_opt}].
One observes that for initial entropies $s_i \lesssim 1$ a
central filling $ \la n_0 \ra \approx 1.5$ is always close
to optimal.

A special situation arises when we approach zero
temperature. As shown in Fig.~\ref{fig_P1_opt}d, the
quantity $s_f/s_i$ changes periodically when decreasing the
trap strength $b/U$. In this regime an additional MI phase
with $\nu=2$ is present in the center of the trap and the
splitting of the entropy between the lower and upper
fermionic band depends dramatically on the value of the
harmonic confinement. In particular, one can find trap
strengths (e.g. $U/b\approx 2000$ in
Fig.~\ref{fig_P1_opt}d) at which the final entropy
approaches zero and only a comparatively small number of
particles is lost ($N_f/N_i=0.9$).  For shallower traps the
$\nu=2$ MI phase starts to collapse. In this regime it
becomes very difficult to find proper thermal states which
match the initial conditions in terms of entropy and
particle number. Finally, the central occupation approaches
one and the cooling protocol leaves the initial state
unchanged.

It is also interesting to study the efficiency of filtering
when acting on the full 3D lattice. In this case only a
small subset of 1D tubes will satisfy the optimal initial
conditions. Using the parameters of Sect. \ref{Init}
($s_i=1$ and $N=2 \ 10^5$) we obtain that $s_f/s_i=0.78$
and $N_f/N_i=0.62$. The relatively large particle loss
results from the high densities in the center of the trap.
This is also the main reason why the protocol performs
worse compared to a 1D tube with $s_i=1$ as shown in Fig.
\ref{fig_P1_opt}b.

\subsection{Experimental realization of continuous filtering}
\label{sec_exp}

Filtering constitutes a fundamental tool in all our cooling
protocols. As shown before \cite{RZ03,Bloch_filter06,Cool1}
filter operations can be realized by coherently
transferring particles between two internal states and then
removing these particles. What we will show below is that
both processes can be combined producing an incoherent
evolution that gives rise to the completely positive map
$F_1$ (\ref{Fn}).

\begin{figure}[h]
  \includegraphics[width=0.7\linewidth]{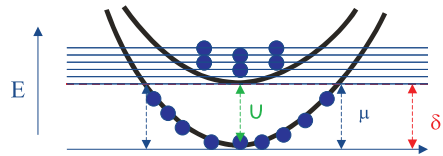}
  \caption{Possible physical realization of incoherent and continuous
    filtering. Atoms with energy $E\geq U$ are resonantly coupled to a
    continuum of untrapped states and thereby removed from the system.}
  \label{fig_levels}
\end{figure}

The experimental procedure for achieving the map $F_M$ is
very simple. We will use two atomic states: one atomic
state shall be confined by an optical lattice, deep in a MI
phase, while the other one will be in a continuum of
untrapped states which are free to escape the lattice. We
will couple the trapped and untrapped states with two Raman
lasers which have a relative detuning of the order of the
interaction energy, $\delta \sim M U$
[Fig.~\ref{fig_levels}]. As long as the coupling is active,
the lasers will depopulate lattice sites which have too
many atoms, $n>M$, while leaving other sites untouched
\footnote{Additionally, particles from the tails may also
be coupled to the continuum and expelled. In order to avoid
that this process leads to an unwanted increase of entropy
in the tails, one can lower the harmonic confinement once
the MI phase has been reached.}.

If the untrapped states are such that they have few atoms
and they are quickly expelled from the trap (for instance
by a magnetic field gradient), these states will behave for
practical purposes like a thermal bath in a vacuum state.
If the coherence time of this bath, physically determined
by the time free atoms spend close to the trapped ones, is
very short compared to the Rabi frequency $\Omega$ of the
Raman transition, we will be allowed to write a master
equation for the trapped atoms. The solution of this
equation converges at large times to the desired filtered
state,  e.g.  $M=1$ for $F_1$.

Conceptually, this mechanism is equivalent to the frequency
knife from evaporative cooling, where atoms containing too
much kinetic energy are expelled from the trap in order to
lower the temperature. In our case, however, it is the
interaction energy we get rid off and, as a side effect, we
make the filling of the lattice more uniform.

Compared to the optimal control scheme in \cite{Cool1}, the
operation of this much simpler method is not very fast.
From the solution of the master equation it follows that
states with occupation $n>1$ decay after a time which is of
the order of the inverse Rabi frequency, $t_1\sim 1/(n
\Omega)$. The main source of errors arises from the
non-resonant coupling of the $n=1$ state with the
reservoir. The probability of a defect (empty site) will
approximately be given by: $p_0\approx \Omega^2/U^2$.
Hence, for $p_0=10^{-4}$ we get an operation time $t_1 \sim
100/U$ which is comparable to the time scale of the
adiabatic scheme \cite{RZ03}.  However, our incoherent
scheme has two big advantages. First, it can be applied
continuously.  Second, it does not put any constraints on
the interaction energies of the two species. This holds
under the assumption that untrapped atoms are expelled so
quickly from the trap, that they do not interact
significantly with the trapped atoms.

\section{Ground state cooling} \label{Sect_gscool}

We have seen that the residual entropy after filtering is
concentrated at the borders of the density distribution.
Particles on these sites are also the only source of energy
excitations because all doubly occupied sites have been
removed. In the following we will propose several protocols
which selectively remove particles at the borders, thereby
bringing the system closer to its ground state.

\subsection{Algorithmic Cooling}
\label{algo-cool}

\begin{figure}[t]
\includegraphics[width=0.7\linewidth]{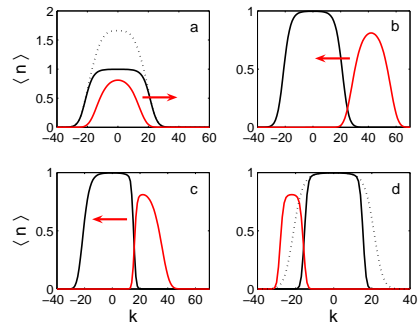}
\caption{Illustration of the algorithmic protocol. The
state is initialized
  with the filter operation $F_2$. (a) Operation $U_{2,0}^{1,1}$: particles
  are transferred to state $|b\ra$ (red). (b) Operation $S_{k_0}$: Lattice
  $|b\ra$ has been shifted $k_0$ sites to the right.  (c) Iteration of the
  sequence: $U_{1,1,0}^{0,0,2}$, $E_c$, $S_{-1}$. If a site is occupied by
  two particles of different species, then both particles are removed.
  Afterwards lattice $|b\ra$ is shifted one site to the left. After $k_s$
  iterations lattice $|b\ra$ is shifted $2 k_0 - k_s$ sites to the left. (d)
  The same sequence as in (c) is applied, with the difference that lattice
  $|b\ra$ is shifted to the right. The final distribution of atoms in state
  $|a\ra$ is sharper compared to the initial distribution (dotted).
  Numerical parameters: $N_i=65$, $s_{2,i}=0.82$ ($s_i=1$), $U/b=300$,
  $k_0$=42, $k_s=20$, $N_f=30.2$, $s_{2,f}=0.19$. After equilibration:
  $s_{2,f}=0.23$ ($s_f=0.31$).}
\label{fig_gs_cool}
\end{figure}

This is a cooling method based on a set of discrete steps
which resemble a quantum information processing protocol.
The steps are the following: (i) We begin with a cloud in
thermal equilibrium in the no-tunneling regime, having two
or less atoms per site, all in internal state $|a\ra$. This
can be ensured with a filtering operation $F_2$. (ii) We
next split the particle distribution into two, with an
operation $U_{2,0}^{1,1}$ [Fig.~\ref{fig_gs_cool}a]. (iii)
The $|a\ra$ and $|b\ra$ atoms are shifted away from each
other until both ensembles barely overlap. We then begin
moving the clouds one against each other, removing in the
process atoms from doubly occupied sites.  Experimentally,
this can be achieved by introducing a third internal level
$|c\ra$ and applying the unitary operation
$U_{1,1,0}^{0,0,2}$ in generalization of (\ref{U_nm}). This
sequence sharpens the density distribution of both clouds
and it is iterated for a small number of steps, of order
$\Delta$ [Fig.~\ref{fig_gs_cool}c].  (iv) The atoms of type
$|b\ra$ are moved again to the other side of the lattice
and a process similar to (iii) is repeated
[Fig.~\ref{fig_gs_cool}d].  (v) Remaining atoms in state
$|b\ra$ can now be removed.

Notice that the final particle distribution will never be
perfectly sharp [Fig.~\ref{fig_gs_cool}d] because this
protocol is intrinsically limited by thermal and quantum
fluctuations in the initial state. Qualitatively, in order
to remove a particle of type $|a\ra$ from the tail, it has
to coincide with a particle of type $|b\ra$ from the other
cloud. Errors arise when particles do not coincide and are
thus proportional to the fluctuations of the density. If we
want to clean about $\Delta$ sites or remove $\Delta$
particles, the errors will be ${\cal O}(\sqrt{\Delta})$. In
the limit $\beta U \gg 1$, we obtain the following scaling
behavior for the final entropy per particle
\begin{equation}
  \label{sf_gscool}s_f \propto \frac{\sqrt{\Delta}}{ \overline{N}} =
  \frac{1}{\sqrt{\overline{N}} ~ \sqrt{\beta \mu}} \propto
  \frac{1}{\sqrt{N_i}},
\end{equation}
where $\overline{N}$ (\ref{N_bar}) is the number of
particles in $|a\ra$ after step (ii) and $N_i$ is the
initial particle number. This simple estimate is already in
good agreement with the exact result in \cite{Cool1}, which
takes also particle losses into account. Our findings show
that the algorithmic protocol becomes more efficient with
increasing initial particle number. This is in contrast to
filtering, for which $s_f$ is independent of $N_i$.

\begin{figure}
  \includegraphics[width=0.5\linewidth]{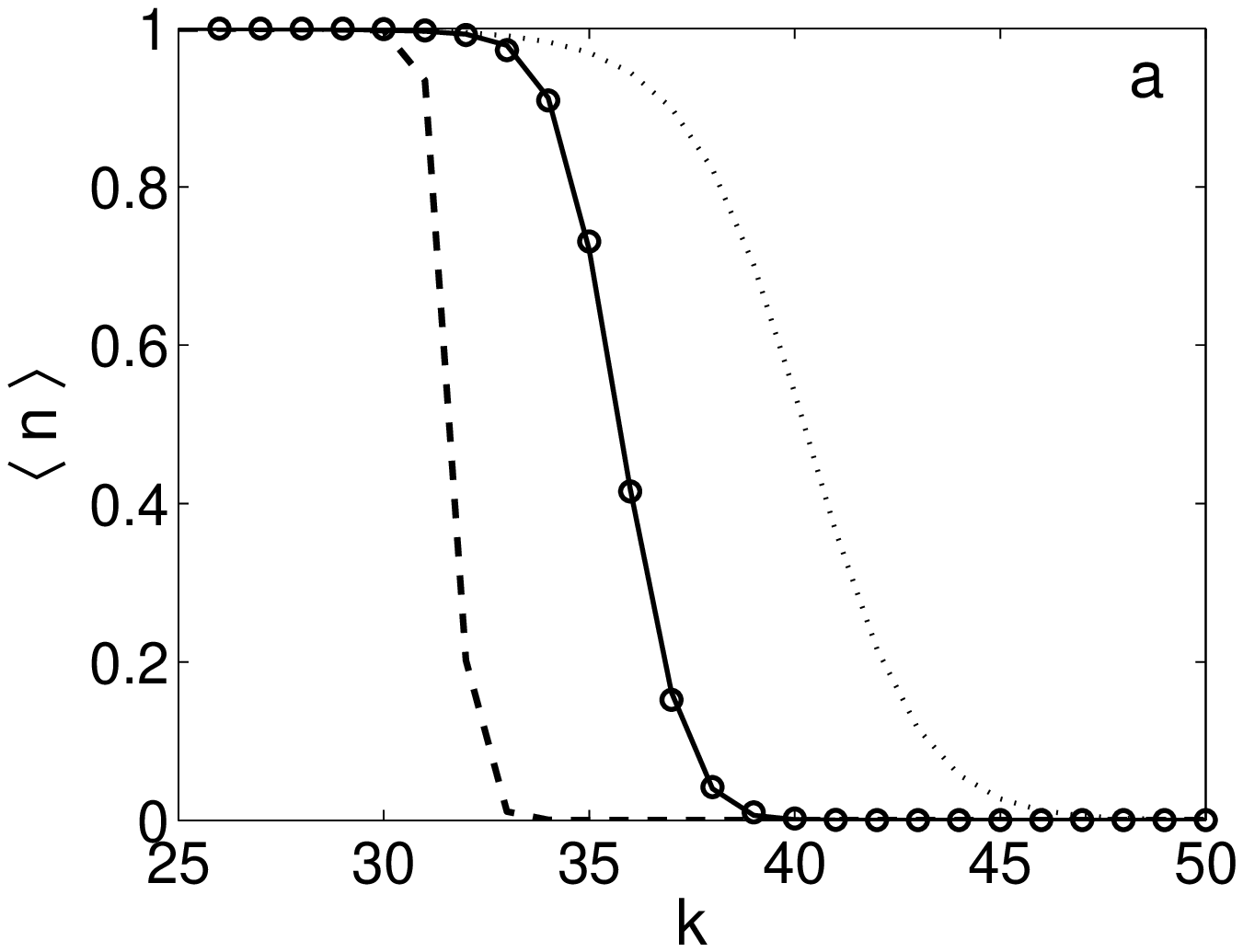}
  \includegraphics[width=0.5\linewidth]{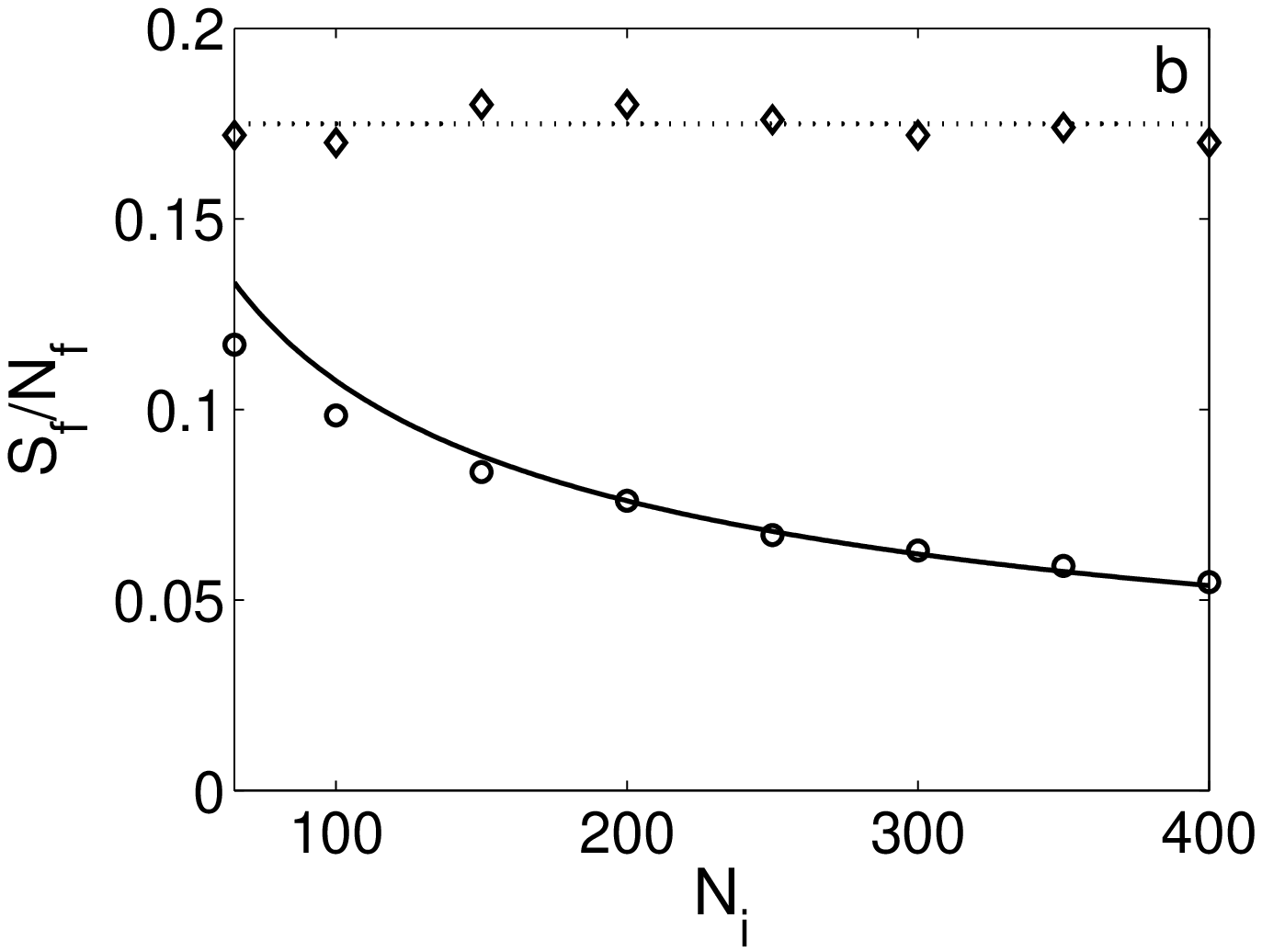}
  \caption{(a) Density distribution after filtering (dotted) and after
    algorithmic cooling (solid) which can be well approximated by a thermal
    distribution (circles). For comparison we consider a variant of the
    algorithmic protocol, which exhibits no classical correlations at all
    times (dashed). Numerical parameters: $N_i=100$, $s_{2,i}=0.42$
    ($s_i=0.50$), $U/b=1500$, $k_0=62$, $k_s=18$, $N_f=71.4$,
    $s_{2,f}=0.0659$; After equilibration: $s_{2,f}=0.0740$ ($s_f=0.0988$);
    Dashed line: $k_s=28$, $N_f=63.3$, $s_f=0.0389$. (b) Final entropy per
    particle $s_f$ as a function of the initial particle number $N_i$ for
    fixed $s_i=0.5$ (circles). For large $N_i$ one obtains a $1/ \sqrt{N_i}$
    dependence (solid). For comparison we plot $s_f$ after filtering
    (diamonds) which is expected to be independent of $N_i$ (dotted).  Small
    variations in $s_f$ can be attributed to the specific choice of the
    harmonic confinement.} \label{fig_gs_cool_sqrtN}
\end{figure}

We have verified numerically the scaling (\ref{sf_gscool})
[Fig.~\ref{fig_gs_cool_sqrtN}b]. The numerical simulation
is by no means trivial because the protocol establishes
classical correlations both among lattice sites and also
among internal states. To reproduce these correlations we
have resorted to representing the classical density
matrices using MPS [Appendix A]. These numerical
simulations show that the final density distribution is
very close to a thermal distribution
[Fig.~\ref{fig_gs_cool_sqrtN}a]. Indeed if we apply one or
few iterations of the protocol, the final state will  still
contain some holes on the tails and be close to thermal
equilibrium, as the computation of the R{\'e}nyi entropy
entropy $s_{2,f}$ (\ref{Renyi}) before and after
equilibration shows [Fig.~\ref{fig_gs_cool_sqrtN}a]. On the
other hand, for a large number of iterations, the final
density matrix will be an incoherent superpositions of
perfect uniform MI states which differ on their length and
position [Similar to Fig.~\ref{fig_P1a_mps}]. These states
are suitable for quantum computation but are far from
equilibrium.

It is clear that we could cool the atoms to the ground
state with $100\%$ efficiency if the density distributions
of $|a\ra$ and $|b\ra$ atoms were perfectly correlated. It
seems also that we can improve the performance in realistic
situations by removing the classical correlations which are
established during the process [Fig.~
\ref{fig_gs_cool_sqrtN}a]. We also increase the performance
if we break the correlations between the atomic species at
the end of the protocol while  leaving the inter--site
correlations untouched. Each iteration of the protocol
would then reduce the total entropy by a factor of
$\sqrt{2}$ \cite{Cool1}. From an experimental point of
view, this means that the algorithm will work better when
using multiple independent clouds to clean each other.
These clouds may come from loading the lattice with atoms
in different internal states, from splitting the lattice
into multiple condensates, or simply by using the clouds
from different 1D tubes to clean each other$\ldots$ Many
other possibilities can be conceived. The protocol can be
further improved by selecting specific particle number
subspaces from the density matrix before restoring thermal
equilibrium (see Sect. \ref{Sect_qrcool} for details).

Note also that we need not get rid of the atomic cloud in
state $|b\ra$. For appropriate initial trap parameters, it
is preferably to move this cloud back to the center and to
pump all atoms back into internal state $|a\ra$. The
resulting state will exhibit a MI shell structure with two
plateaus at densities $n=1$ and $n=2$.  This is a simple
way to engineer ground states of tighter traps. Finally,
let us remark that this protocol will only correct
inhomogeneities along one direction. In a real experiment
the protocol should be repeated along the other three
directions.

\subsection{Non-algorithmic cooling schemes} \label{Alternative}

We have seen that algorithmic schemes perform much better
than simple filtering. Here, we propose two non-algorithmic
ground state cooling schemes which combine filtering with
other quantum operations available in optical lattices.

\subsubsection{Sequential filtering}

The basic idea of this protocol is to first transfer
particles from the tails of the cloud to the center of the
trap, to form doubly occupied sites, and then to remove
these particles with some filtering operation. To be more
precise, one has to iterate the following sequence of
operations: (i) Allow for some tunnelling while the trap is
adjusted in order to reach a central occupation of $\la n
\ra_0 \approx 1.5$. (ii) Suppress tunnelling and perform
the filtering. (iii) Slightly open the trap so that final
distribution resembles a thermal one (See
Sect.~\ref{Sect_filtering}). These steps are repeated
sequentially until the entropy per particle converges.

This protocol has been analyzed in detail in \cite{Cool1}
and the results can be summarized as follows. If one allows
for equilibration then a state asymptotically close to the
ground state can be reached after only a few filtering
cycles. Without equilibration the final entropy is limited
by the initial probability of finding a defect in the
center of the trap.

\begin{figure}[t]
  \includegraphics[width=0.5\linewidth]{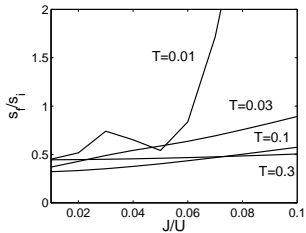}
  \caption{Cooling efficiency $s_{2,f}/s_{2,i}$, in terms of the R\'enyi
    entropy, of filtering $F_1$ as a function of the hopping rate $J/U$. For
    the numerical computation in terms of MPS we use a lattice of length
    $L=27$, which typically contains $N \approx 20$ particles. The trap
    strength is chosen to be $b/U=1/65$ and $\mu=U$, yielding $\la n \ra_0
    \approx 1.5$.}
  \label{fig_dynamic_3}
\end{figure}

Alternatively, one can think of an implementation of
sequential filtering that operates at fixed but non-zero
hopping rate. Such a protocol would clearly profit from the
fact that adiabatic changes of the lattice potential, which
lead to additional heating, are not required. A key
ingredient is the generation of doubly occupied sites in
the center. One has to fix the hopping rate at a value, at
which there exists a coupling between doubly occupied sites
at the center and singly occupied sites at the borders. Our
analysis of the BHM in terms of an effective two-band
fermionic model \cite{Cool1} shows, however, that this
occurs typically only deep in the SF regime for $J/U\gtrsim
0.5$.

We have studied the sequential filtering at fixed $J$
numerically, computing the cooling efficiency as a function
of the hopping rate. The results are shown in
Fig.~\ref{fig_dynamic_3}.  For high temperatures the ratio
$s_f/s_i$ is rather independent of $J$ and we achieve some
cooling. For very small temperatures, $k T \lesssim 0.01
U$, it changes dramatically with $J$. This is a clear
signature of the quantum phase transition, which is
expected to occur at $J_c \approx 0.085 U$ in the
thermodynamic limit. While for $J < J_c$ the state is well
described in terms of independent wells and filtering works
very efficiently, for $J > J_c$ particles are delocalized
over the lattice, and filtering causes heating rather than
cooling. Summing up, since we require a significant value
of the hopping, $J/U\sim 0.5$, this second variant of
sequential filtering can only be used as an initial step,
when temperatures are still comparatively high.

\subsubsection{Filtering combined with frequency knife}

Particles located at the tails of the density distribution
can also be removed with a method similar to evaporative
cooling or a frequency knife. The idea is to make use of
inhomogeneous on-site energies and to choose the detuning
$\delta$ of a radiation field in such a way that only atoms
located at specific lattice sites are resonantly coupled to
another internal state. Using a magnetic field gradient it
has been demonstrated that individual lattice sites can be
resolved within an uncertainty of about five sites
\cite{Meschede04}. In our case the spatial dependence of
the on-site energies is naturally provided by the harmonic
trapping potential. However, in order to make use of this
inhomogeneity one has to ensure that the atoms are coupled
to an internal state which responds differently to the
ac-Stark shift induced by the lattice laser beams.
Experimentally, this can, for instance,  be achieved with a
setup similar to the one for the creation of spin-dependent
lattices \cite{OL_ent,Bloch03,OL_spin_99}. There, one
atomic species is trapped exclusively by $\sigma^-$
polarized laser light, whereas the other species is trapped
predominantly by $\sigma^+$ polarized  light. Hence, an
optically untrapped internal state can simply be realized
by using only $\sigma^+$ polarized laser light for creating
the optical lattice.

Let us now estimate the requirements on the the Rabi
frequency $\Omega$ of the transition depending on the trap
strength $b$. For simplicity, we consider two internal
states, and a configuration for which the excited state
exhibits zero on-site energy at each lattice site. We are
interested on removing particles which are typically
located at a distance $k_\mu$ from the center of the trap.
Hence, for resonant coupling of these atoms we must choose
the detuning to be $|\delta|=b k_\mu^2=\mu$. Particles at
sites $k_\mu + \Delta k$ feel an effective detuning
$\delta_{eff}(\Delta k)\approx b \overline{N} \Delta k$,
where $\overline{N}=2 k_\mu$ denotes the particle number
(\ref{N_bar}). The probability that a particle at site
$k_\mu + \Delta k$ is transferred to the excited state is
then given by:
\begin{equation}
  P_e(\Delta k)=\frac{ \Omega^2}{ \Omega^2+ \delta_{eff}(\Delta k)^2}.
\end{equation}
In order to locally address the site $k_\mu$ reasonably
well one has to demand: $\Omega \lesssim b \overline{N}$.
For typical harmonic trapping frequencies in the MI regime,
$\omega_{ho}=\sqrt{8 b/(\hbar m
  \lambda^2)}\gtrsim 2 \pi \times 65$ Hz \cite{Bloch02}, and
$\overline{N}=50$ this translates into $\Omega \lesssim1$
kHz.

Experimentally it should therefore be feasible to resolve
individual sites with an uncertainty of a few lattice sites
and thus to sharpen the density profile within this
uncertainty. Note that this scheme clearly profits from a
large number of particles per tube. Another advantage is
that it preserves the spherical symmetry of the density
distribution in a 3D lattice.

Finally let us remark that this method can easily be
incorporated in the continuous filtering scheme
[Fig.~\ref{fig_levels}] proposed in
Sect.~\ref{Sect_filtering}, since there the atoms are also
coupled to untrapped states.

\section{Algorithmic cooling of defects in quantum registers}
\label{Sect_qrcool}

A perfect one-dimensional quantum register is a connected
array of commensurately filled lattice sites.  For most
purposes the filling factor is $\nu=1$. This state appears
naturally as the unique ground state of the BHM in the
no-tunnelling limit and in the presence of harmonic
confinement. Hence, any efficient ground state cooling
protocol will produce a good quantum register. Moreover, we
have seen in the previous section, that algorithmic
protocols can be used to create an \emph{ensemble} of
quantum registers rather than a unique one (see also
\cite{VC04,Cool1}). In this section we will propose and
analyze two alternative algorithms for the creation of a
quantum register ensemble. As compared to previous
proposals \cite{VC04,Cool1} these schemes require only a
small number of operations, which makes them more appealing
for experimental implementation. The first protocol
produces registers with filling $\nu=1$, whereas the second
protocol is optimized for filling $\nu=2$. In addition, we
propose how to transform atoms at the endpoints of each
register into "pointer" atoms, which then enables
addressing individual lattice sites. This can be used, for
instance, to create registers of equal length. Even more
important it  offers the opportunity to perform ensemble
quantum computation in this system.

We start from a rather cold cloud, $\beta U \gg 1$, which
has been subject to fundamental filtering operations $F_M$
(\ref{Fn}). The result is an almost perfect MI in the
center of the trap with some residual defects (or holes)
which are predominantly localized at the borders [See
Fig.~\ref{fig_P1}]. Since we operate solely in the
no-tunnelling regime these defects cannot redistribute nor
evaporate. Our goal is to remove all these defects and we
will achieve it by applying nearest-neighbor quantum gates
which simulate inelastic collisions between particles and
holes. Simply put, whenever a particle sits next to a hole,
the particle will be annihilated. This process is analogous
to spin flips in ferromagnets and the formation of domains
of equal magnetization. Thus, these algorithmic schemes can
also be understood as controlled equilibration and cooling
of defects.

For all protocols that will be proposed below, the
resulting state is a mixture of perfect (up to defects in
the central MI phase) quantum registers, which differ only
in their length and lateral position. The entropy of this
state will be of the order $S_f\sim \log_2 \Delta$, where
the parameter $\Delta$ (\ref{Delta}) quantifies the
translational uncertainty in the initial density
distribution. Note, however, even though the final entropy
is very small, these protocols are typically not suited for
ground state cooling. The reason is that the final state is
far from thermal equilibrium. Numerically we find that the
value of the entropy after equilibration is comparable to
the value before invoking the protocol. However, if one
makes use of the pointer atoms  to select only  registers
with a specific length then these protocols indeed lead to
cooling.

\subsection{Protocol 1}

\begin{figure}
  \includegraphics[width=0.8\linewidth]{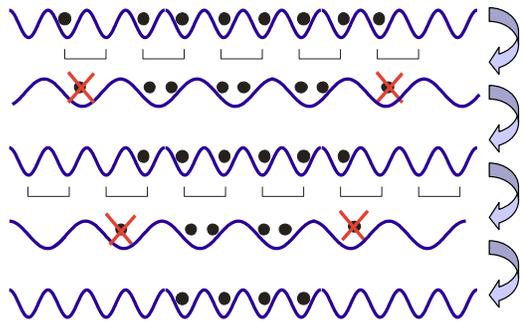}
  \caption{Illustration of protocol 1 for the creation of quantum registers.
    After merging neighboring lattice sites, one empties single occupied
    sites. The algorithm stops, when all particles in the tails have been
    removed and only the central MI phase remains.}
  \label{fig_P1a_lattice}
\end{figure}

The operation sequence is illustrated in
Fig.~\ref{fig_P1a_lattice}. We first merge oddly aligned
pairs of sites using a superlattice
\cite{Calarco,Calarco_1} and empty sites with only one
atom. The sites are again split and the original lattice is
restored. This operation is then repeated several times,
alternating between even and oddly aligned sites, until the
\emph{total} entropy reaches a minimum.

\begin{figure}
  \includegraphics[width=0.48\linewidth]{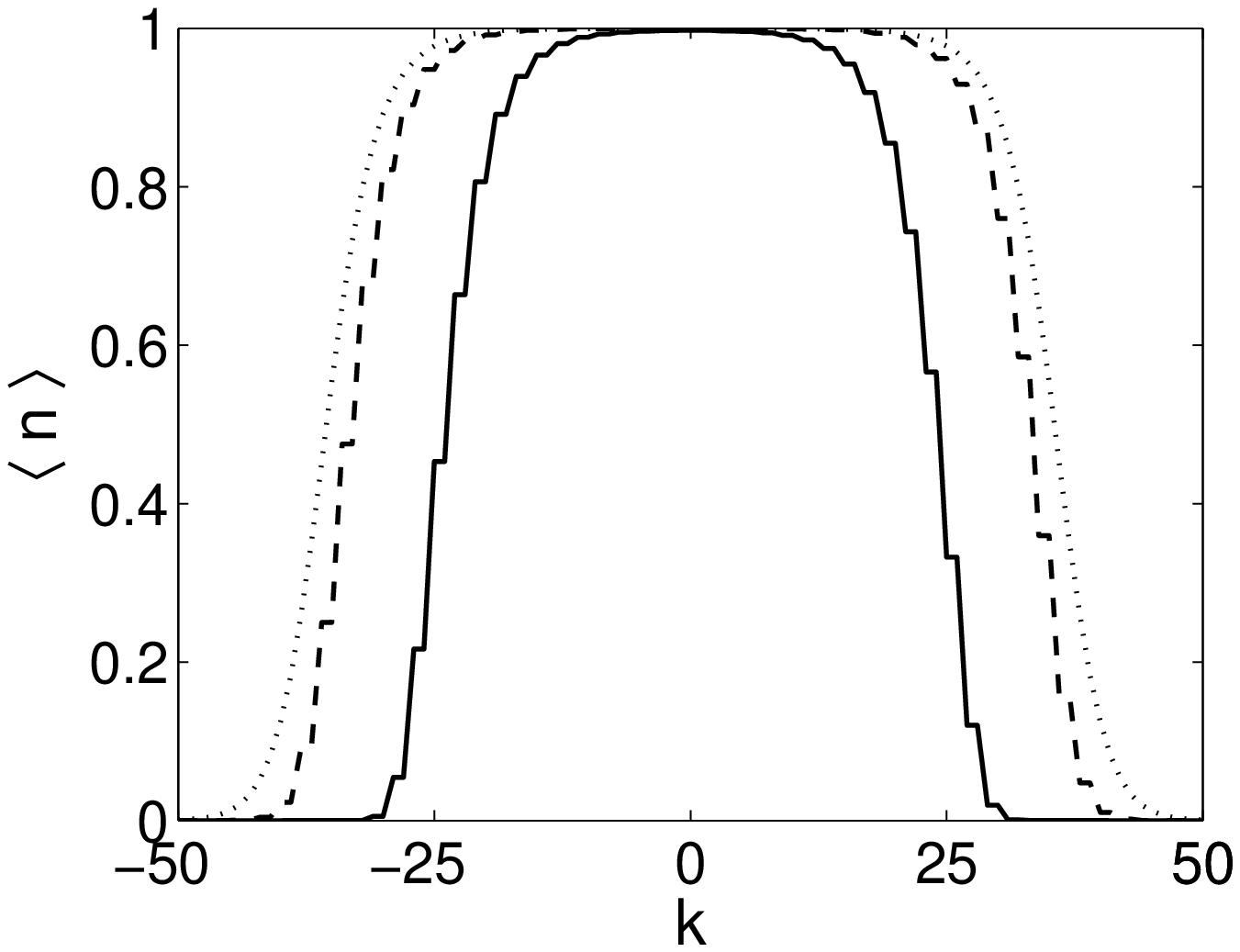}
    \includegraphics[width=0.48\linewidth]{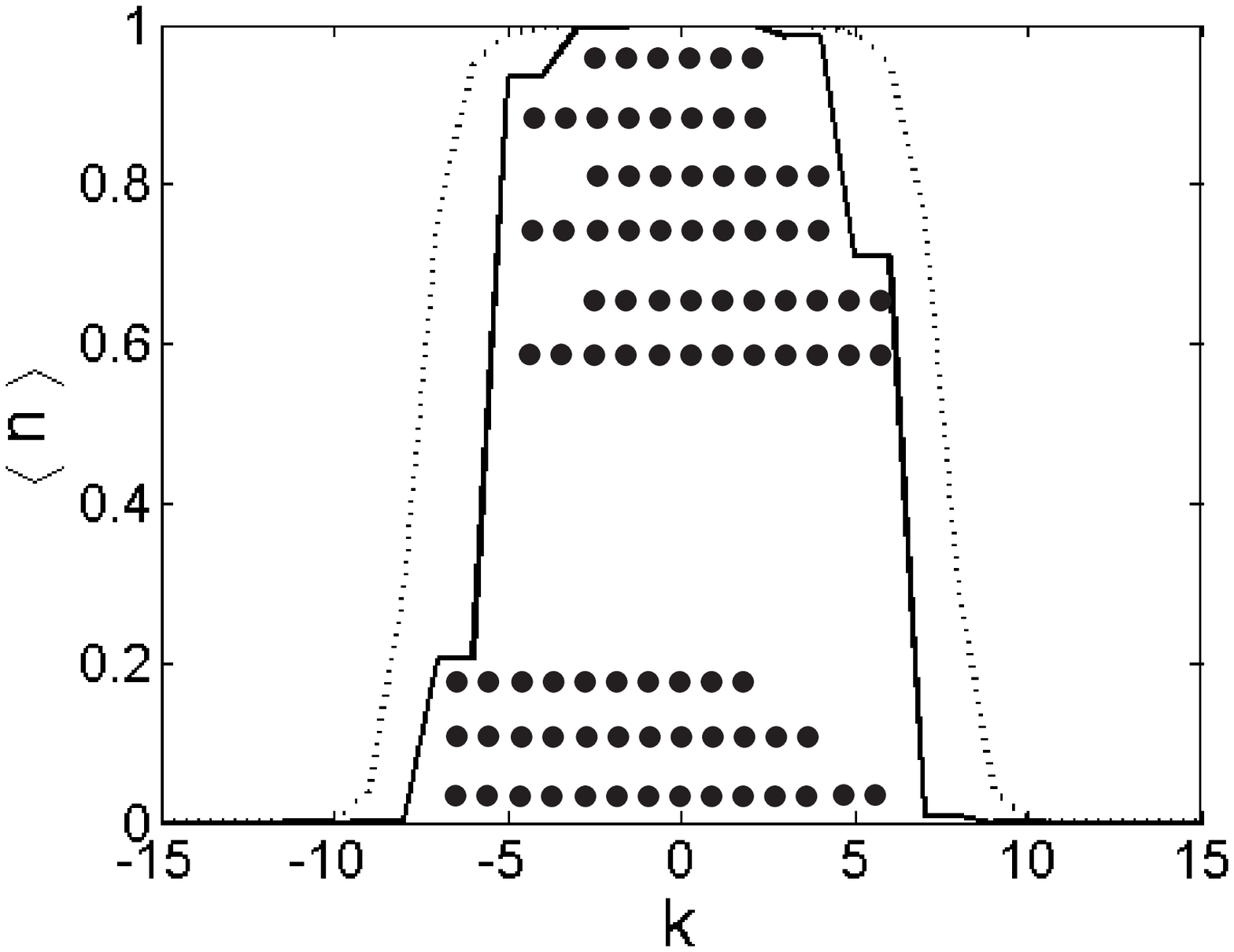}
  \caption{Left: Particle distribution after different steps of the protocol:
    after filtering $F_1$ (dotted), after one iteration (dashed), after
    eight iterations (solid). Parameters of the initial state: $N=100$,
    $s=0.7$ ($s_2=0.56$), $U/b$=1000, $\beta U=5.8$.  Parameters after filtering:
    $\overline{N}=70.5$, $s=0.33$ ($S_2=17.5$, $s_2=0.25$). Parameters after iteration
    1: $N=65.5$, $S_2=7.89$, $s_2=0.12$; iteration 2: $N=60.8$, $S_2=5.95$,
    $s_2=0.098$; iteration 7: $N=48.8$, $S_2=5.297$, $s_2=0.108$; iteration
    8: $N=46.8$, $S_2=5.296$, $s_2=0.113$. Stopping the algorithm at
    iteration 3 gives a minimum in the entropy \emph{per} particle:
    $s_2=0.095$. Upon reaching thermal equilibrium this value has increased
    to $s_2=0.265$ ($s=0.36$) which is comparable to the value after filtering. However, when
     selecting
    the subspace containing only registers of length $N=48$ then the protocol leads to cooling.
    The entropy per particle in thermal equilibrium will then be given by $s=0.17$. Right: The step
    like structure of the density distribution allows one to
    deduce the states which contribute significantly to the final density
    matrix. The states
    can be classified according to their particle number and their lateral
    position. }
  \label{fig_P1a_mps}
\end{figure}

We have analyzed the performance of this protocol under
realistic conditions using the MPS description to compute
the at most classically correlated density matrix [Appendix
A]. In Fig.~\ref{fig_P1a_mps} we plot the typical density
distribution after different steps of the protocol. A
single step changes very little the density but decreases
dramatically the entropy (From $S_2=17.5$ to $S_2=7.9$).
Indeed, the first steps account for the elimination of most
defects. After a few iterations the value of the total
entropy starts to saturate because the density matrix has
collapsed to a classical ensemble of (almost) defect-free
quantum registers [Fig.~\ref{fig_P1a_mps}]. As a
consequence of the protocol, the number of atoms per
register must be even, which leads to steps in the final
density profile [Fig.~\ref{fig_P1a_mps}].

Clearly the fix point of this protocol would be a state
with no particles at all. Let us now estimate how many
iterations $M$ of the protocol are required to reach a
state with reasonable particle number and tolerable defect
probability. First, we have to point out that there are two
sources of defects: (i) holes in the central MI phase and
(ii)  particles at the borders which are disconnected from
the central MI phase and which have not yet been erased by
the protocol. The probability for defects of the first kind
is negligible in the limit of low temperatures $\beta U\gg
1$ [see Sect. \ref{Sect_initstate}]. Defects of kind (ii)
can be assessed by the following observation: In order to
erase a connected array of $M$ particles, which is
separated by at least one empty site from a central MI
state, one requires exactly $M/2$ iterations of the
protocol. The probability of finding defects after $M$
iterations is then given by the probability of having
 states  with array size larger than $M$ in the
initial density matrix. This probability can easily be
computed numerically exact from the distribution
(\ref{nf}). The derivation of closed expressions is however
difficult. Nevertheless, one can get a good estimate for
the optimal number of iterations based on the following
argument. The characteristic tail width of the initial
particle distribution (\ref{nf}) is $\Delta$ (\ref{Delta}).
Hence, the occurrence of arrays of size $M=2 \Delta$ is
already very unlikely. This implies that after roughly
$\Delta$ iterations of protocol 1 we expect to have
registers with negligible defect probability. This can be
illustrated with an example. For the initial state in Fig.
\ref{fig_P1a_mps} we have $\Delta = 10$ and the relative
difference in the total entropy after the seventh and
eighth  iteration has reduced  to $10^{-4}$, which implies
a defect probability of the same order of magnitude. The
typical size of the registers after $\Delta$ iterations is
roughly $\overline{N}- 2 \Delta$, where  $\overline{N}$
(\ref{N_bar}) is the characteristic size of the initial MI
state. This  implies  $\overline{N} \gg \Delta$,  otherwise
no particles remain in the system. Since $\Delta \approx
\overline{N}/\beta U$, one has to require  the low
temperature regime $\beta U \gg 1$.

As mentioned above, the algorithmic ground state cooling
protocol of the previous section can also be used for the
creation of quantum registers. However, the ground state
cooling algorithm involves $\mathcal{O}(\overline{N})$
operations as compared to $\mathcal{O}(\Delta)$ operations
of the current protocol.

Note also that the example presented in Fig.
\ref{fig_P1a_mps} indicates that this protocol does not
lead to cooling. After restoring thermal equilibrium, the
entropy has reached a value comparable to the one after
filtering. We have repeated this analysis for various
initial conditions and our results confirm this
observation. However, if one selects a particle number
subspace with $N$ being larger than the mean value than our
protocol can indeed be used for cooling. According to the
data in Fig. \ref{fig_P1a_mps} the entropy per particle can
be reduced by roughly  $50\%$ as compared to filtering.
\subsection{Protocol 2}

We start from a state which contains only empty or doubly
occupied sites. This can be achieved by applying filtering
operation $F_2$, followed by $U_{1,0}^{0,1}$ and $E_b$.
\begin{figure}[t]
  \includegraphics[width=0.7\linewidth]{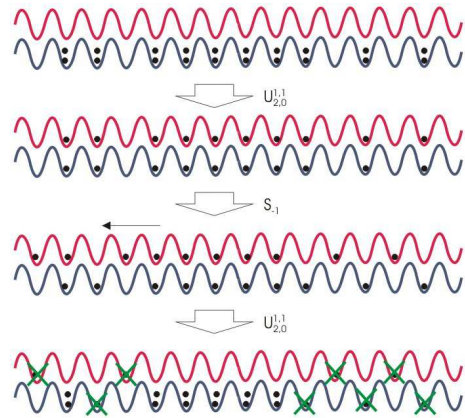}
  \caption{Illustration of protocol 2 for the creation of quantum registers.
    Spin-dependent lattice shifts allow to remove defects in a correlated
    way. }
  \label{fig_P2a_lattice}
\end{figure}
The protocol is depicted in Fig.~\ref{fig_P2a_lattice}. We
begin with transferring one particle per site to state
$|b\ra$. Then the $|b\ra$--lattice is shifted one site to
the left and the same operation as before is performed.
Afterwards one empties single occupied sites.  This
procedure allows to remove defects in a correlated way.
Occupied sites which have an unoccupied site to the right
are emptied.  Since the probability of finding particles in
the center is close to one, the central MI is preserved
except for losses at the right border.  This procedure is
repeated until all particle disconnected from the central
MI phase  become annihilated and only perfect MI phases in
the center remain. One step of the protocol can be
summarized in the following sequence of operations:
$U_{2,0}^{1,1}$, $S_1$, $U_{2,0}^{1,1}$, $E_b$,
$U_{1,0}^{0,1}$, $E_b$. Following the discussion of
protocol 1, this sequence has to be applied approximately
$2 \Delta$ times, where $\Delta$ denotes again the
characteristic tail width of the initial particle
distribution. The factor two stems from the fact that at
each step of the protocol the size of "particle islands" in
the tails, as well as the central MI phase, is reduced only
by one as compared to two in protocol 1. The final density
matrix looks very similar to the one after protocol 1, with
the difference that also registers with odd number of atoms
appear.

\subsection{Pointer atoms and register length control}

Given an ensemble of quantum registers we will now show how
to create pointer atoms. To be more precise, our goal is to
selectively transfer the two end atoms of each register to
a different internal level. These pointer atoms enable
single--site addressing which can used to engineer
registers of specific length.
\subsubsection{Creation of pointer atoms}

 Our scheme relies on the same set of operations that is
used in current experiments for entangling atoms located at
different lattice sites \cite{Bloch03}. We consider quantum
registers with one atom per site. Initially all atoms are
in internal state $|a\ra$. A Hadamard transformation puts
the atoms in the coherent superposition state $(|a\ra+
|b\ra)/ \sqrt{2}$. One then shifts the $|b\ra$--lattice one
site to the right and waits an appropriate time until the
on--site interaction between species $|a\ra$ and $|b\ra$
yields a collisional $\pi$--phase. This means that on each
site the state $|a\ra |b\ra$ is transformed into $-|a\ra
|b\ra$. After two lattice shifts to the left one waits
again until a $\pi$--phase has built up. Then the lattice
is shifted back to its original position and a second
Hadamard operation is performed. The resulting state is a
again a product state but the end atoms have been promoted
to level $|b\ra$. These atoms can be considered as pointer
atoms because they mark the beginning (and the end) of each
register and can thus be used to access every site within
the register in a deterministic way. In practice, only one
pointer atom is needed. The second pointer atom, e.g. the
one on the left, can easily be removed by applying the
following operation sequence: $S_{-1}$, $U_{1,1}^{2,0}$,
$E_ b$, $U_{2,0}^{1,1}$.

\subsubsection{Manipulation of register length}
The pointer atoms can now be used to create an ensemble of
registers of fixed length, a feature which is desired for
quantum computation and which can even lead to  cooling. We
first present a protocol that requires only two internal
states of the atoms. Then we show that the algorithm can be
simplified considerably when a third internal level is
included. In both cases we start from a situation where all
registers have one pointer atom in state $|b\ra$ which is
located at the right most occupied site of the register.

{\bf Protocol based on two internal states: } The central
idea is to remove atoms selectively from the system by
transferring them to the pointer level $| b \ra $. We first
show how to discard all registers which are shorter than a
desired length  $M$. We start with the sequence:
$U_{1,1}^{0,2}$, $S_{-M}$, $U_{0,2}^{2,0}$, $E_ b$. This
ensures that registers of length $N \geq M$ are protected
against further modifications, because their pointer atoms
have been removed. Next we promote atoms in doubly occupied
sites to the pointer level and shift this (two--atom)
pointer one site to the right. If the pointer hits an
occupied site then one atom in $|a\ra$ will be removed.
Iteration of this process  removes all atoms on the right
of the current pointer position.   To be more precise, one
has to iterate $M-1$ times the sequence: $U_{2,0}^{0,2}$,
$S_1$, $U_{0,2}^{2,0}$, $U_{1,2}^{2,1}$, $E_b$. In a last
step the remaining doubly occupied sites are emptied via
$U_{2,0}^{0,2}$, $E_b$. After creating new pointer atoms
the minimal register length is given by $M'=M-2$. Again we
keep only the pointer atom on the right side. Let us now
show how to shorten registers of length $N>M'$ to length
$M'$. We first apply the sequence: $S_{-1}$,
$U_{1,1}^{0,2}$, $S_{-M'}$, $U_{0,2}^{1,1}$, $S_{-1}$. For
the target register of length $M'$ this merely transfers
one atom from the right end of the register to the left
end. For larger registers one obtains a  two--atom pointer
that can now be used to select and discard  all atoms
located left of the pointer. This can be accomplished by
iterating the sequence: $U_{1,2}^{2,1}$, $E_b$,
$U_{2,0}^{0,2}$, $S_{-1}$ until all registers with
appreciable weight in the density matrix have been
shortened to the desired size.

{\bf Protocol based on three internal states: } This
protocol is based on the operation $U_{1,1,0}^{0,1,1}$,
which transfers an atom from $|a\ra$ to $|c\ra$, given that
a pointer atom is present. This way one can use the pointer
as a "marker", which allows one to produce registers of
desired length $M$ in a very simple manner. One first marks
an array of $M$ atoms and then discards all atoms which
have remained in level $|a\ra$. The algorithm can be
summarized as follows: apply the sequence $S_{-(M-1)},
U_{1,1}^{2,0}, E_b, U_{2,0}^{1,1}, U_{1,1,0}^{0,1,1}$;
iterate $M-1$ times the sequence $ S_1, U_{1,1,0}^{0,1,1}$;
finally apply the operation $E_a$.

\section{Conclusion}

We have proposed various ground state cooling schemes that
allow to reduce the temperature in current optical lattice
setups considerably. In particular, we have presented
algorithmic cooling schemes which combine filtering with
concepts inspired by ensemble quantum computation. We have
shown that these algorithmic protocols can be designed both
for ground state cooling and the creation of defect-free
quantum registers. We have tested and analyzed our
protocols under realistic experimental conditions, with
special emphasis on the presence of a harmonic trap. We
found that the residual entropy of all our protocols
depends crucially on one parameter which accounts for the
translational uncertainty of the initial particle
distribution.

A special virtue of our schemes is, that they rely on
general concepts which can easily be adapted to different
experimental situations.  For instance, little
modifications ensure that our protocols can be applied both
to bosonic and fermionic systems. A second advantage of our
protocols is, that they are designed for the no--tunnelling
regime and hence do not necessarily require equilibration
processes induced by particle hopping and elastic
collisions. Hence, they allow to approach the ground state
of the no-tunnelling regime in very short times.

In this sense, the collection of cooling schemes presented
in this article can be considered as a toolbox which is
tailored for cooling atoms in optical lattice setups. The
tool (or combination of tools) which is best suited for a
given purpose, can be chosen according to the
characteristic features of a specific experimental setup.
For instance, in the case of large systems at high
temperatures, one can think of combining filtering with the
frequency knife method which is then followed by the
algorithmic ground state cooling protocol. Or ground state
cooling can be combined with adiabatic transformation of
the Hamiltonian so as to produce ground states of models
different from the simple Bose-Hubbard considered here. And
one should also keep in mind that a 3D lattice structure
offers a large variety of possibilities, which have not
been fully explored yet.

The methods introduced here greatly enhance the potential
of optical lattice setups for future applications and might
pave the way to the experimental realization of quantum
simulation and quantum computation in this system. We also
hope that the concepts introduced in this work might
trigger further research in the direction of ground state
cooling in optical lattices.

\section{Acknowledgements}

This work was supported in part by the EU IST projects
QUPRODIS and COVAQIAL , the DFG (SFB 631) and the
"Kompetenznetzwerk Quanteninformationsverarbeitung der
Bayerischen Staatsregierung".

\begin{appendix}

\section{Numerical description of classical density matrices in terms of MPS}

Our algorithmic protocols establish classical correlations
between different lattice sites, when applied to thermal
states in the no-tunnelling regime. Hence, a description in
terms of independent wells (as in (\ref{rho_ind})) is no
longer adequate. Therefore we refer to a representation in
terms of matrix product states (MPS)
\cite{Fannes,RomerPRL}.

To be more precise, for a 1D lattice of length $L$ we want
to map a classical  density matrix of the form
\begin{equation}
  \rho=\sum_{\{i\}} \rho^{i_1 \ldots i_L} |i_1 \ldots i_L\ra \la i_1 \ldots i_L |
\end{equation}
onto a pure state in MPS form:
\begin{equation}
  \label{MPS}| \psi \ra =
  \sum_{\{i\}} A_1^{i_1} A_2^{i_2} \ldots A_L^{i_L} |i_1 \ldots i_L \ra.
\end{equation}
Here, $A^{i_k}$ denote matrices of dimension $D \times D$
and \mbox{$i_k
 \in \{0, \ldots, d-1 \}$} is the occupation number of site $k$. The matrices at
the endpoints, $A_1^{i_1}$ and $A_L^{i_L}$, are $1 \times
D$ and $D \times 1$ vectors, respectively. The mapping from
$\rho$ to $|\psi \ra$ can easily be accomplished by
setting: $\rho^{i_1
  \ldots i_L}=A_1^{i_1} A_2^{i_2} \ldots A_L^{i_L}$.

  Expectation values for
operators of the form $\hat O = \hat O_1 \otimes \hat O_2
\ldots \otimes \hat O_L$ are calculated according to the
relation
\begin{equation}
  \la \hat O\ra = \tr \ \hat O \rho =\prod_{k=1}^L
  \sum_{i_k=1}^d \la i_k | \hat O_k | i_k \ra A_k^{i_k}.
\end{equation}

Local operations on $\rho$, like unitary operations
$U_{n,m}^{n',m'}$ (\ref{U_nm}) or filter operations
(\ref{Fn}), amount to transformations of the local matrices
$A_k^{i_k}$ in the MPS picture.  For illustration, let us
consider a completely positive map $\mathcal{C}$ which acts
on a local state $\rho_k$ at site $k$. In Kraus
representation one can write
\begin{equation}
  \mathcal{C}(\rho_k)=
  \sum_\alpha E_\alpha \rho_k E_\alpha^\dagger,
\end{equation}
with the Kraus operators $E_\alpha$ satisfying the
completeness relation $\sum_\alpha E_\alpha^\dagger
E_\alpha = \one$. The MPS matrices transform according to
\begin{equation}
  \tilde A_k^{i_k}= \sum_\alpha \sum_{j_k}
  |E_\alpha^{i_k, j_k}|^2 A_k^{j_k}.
\end{equation}

Non-local operations that involve more than one lattice
site are more complicated to implement. As an example, let
us consider the most complicated case, which occurs in our
protocols: a spin-dependent lattice shift $S_{-1}$ which
shifts the lattice $|b\ra$ one site to the left. For this
we have to consider two species of atoms. In generalization
of (\ref{MPS}), the MPS matrices $A_k^{i_k,j_k}$ are now
labelled with two physical indices, $i_k$ and $j_k$,
referring to states $|a \ra$ and $|b\ra$, respectively.  It
will turn out to be convenient to rewrite these matrices in
tensor form: $(A_k^{i_k,j_k})_{\alpha, \beta}=A_k(\alpha,
\beta, i_k, j_k)$, with $\alpha, \beta = 1 \ldots D$.

To begin with let us consider only the first two lattice
sites. We start out with tracing over subsystem $|b\ra$ at
the first site: $B_1(\alpha_1, \beta_1, i_1)=\sum_{j_1}
A_1(\alpha_1, \beta_1, i_1, j1)$, because we want to omit
sites which pass the system boundaries after shifting.  The
induced error is negligible, given that sites at the
boundaries are not populated. We then multiply the matrix
of the second site and perform the lattice shift: $
\Theta(\alpha_1,i_1,\beta_2,i_2,j_2)=
B_1(\alpha_1,\beta,i_1) A_2(\beta, \beta_2,i_2,j_2)
\rightarrow \Theta(\alpha_1,i_1,j_2,\beta_2,i_2)$, where we
use Einstein summation convention. After relabelling $j_2
\rightarrow j_1$ we perform a singular value decomposition:
\begin{eqnarray}
  &&\Theta(\alpha_1,i_1,j_1,\beta_2,i_2)= \nonumber \\
  && U((\alpha_1,i_1,j_1), \gamma) \ \Sigma(\gamma, \gamma') \
  W^\dagger(\gamma',(\beta_2,i_2)),
\end{eqnarray}
with $\Sigma$ being a diagonal square matrix and
$\gamma,\gamma'=1\ldots \tilde D \leq d D$. The new MPS
matrix for site $1$ can then be defined as:
\begin{equation}
  \left(\tilde{A}_1^{i_1,j_1}\right)_{\alpha_1, \beta_1}:=
  U(\alpha_1,i_1,j_1, \gamma) \ \Sigma^{1/2}(\gamma, \beta_1).
\end{equation}
Accordingly, we define
\begin{equation}
  B_2(\alpha_2, \beta_2, i_2) :=
  \Sigma^{1/2}(\alpha_2, \gamma) \ W^\dagger(\gamma,\beta_2,i_2),
\end{equation}
and iterate this scheme until the end of the lattice. Note,
that the dimension $D$ of the MPS matrices can, in
principle, increase exponentially with the number of
lattice shifts. If $D$ becomes larger than a desired value
$D_{max}$ one has to resort to a truncation method similar
to the one proposed for mixed quantum states
\cite{DMRGmixed}. However, in practice, we find that $D$
increases only linearly with the number of lattice shifts,
given that each shift is followed by a filter operation.
Thus, with our method we can fairly easy simulate protocols
exactly, i.e. without truncation, which involve up to $100$
lattice shifts on lattices with up to $L=500$ sites.
\end{appendix}

%\bibliography{bibliography_Cool}

\end{document}